\documentclass[aps,prb,showpacs,twocolumn,superscriptaddress,floatfix]{revtex4-1}

\usepackage{graphicx}  

\usepackage{amsmath,amssymb,bm}   

\usepackage{dcolumn}

\begin{document}


\title{The electronic band structure and optical properties of boron arsenide}

\author{J. Buckeridge} 
\email{j.buckeridge@ucl.ac.uk}
\affiliation{University College London, Department of Chemistry, 20
  Gordon Street, London WC1H 0AJ, United Kingdom}
\author{D. O. Scanlon}
\affiliation{University College London, Department of Chemistry, 20
  Gordon Street, London WC1H 0AJ, United Kingdom}
\affiliation{Thomas Young Centre, University College London, Gower
  Street, London WC1E 6BT, United Kingdom}
\affiliation{Diamond Light Source Ltd., Diamond House, Harwell Science
  and Innovation Campus, Didcot, Oxfordshire OX11 0DE, United Kingdom}

\begin{abstract}

We compute the electronic band structure and optical properties of boron arsenide using the relativistic quasiparticle self-consistent $GW$ approach, including electron-hole interactions through solution of the Bethe-Salpeter equation. We also calculate its electronic and optical properties using standard and hybrid density functional theory. We demonstrate that the inclusion of self-consistency and vertex corrections provides substantial improvement in the calculated band features, in particular when comparing our results to previous calculations using the single-shot $GW$ approach and various DFT methods, from which a considerable scatter in the calculated indirect and direct band gaps has been observed. We find that BAs has an indirect gap of 1.674 eV and a direct gap of 3.990 eV, consistent with experiment and other comparable computational studies. Hybrid DFT reproduces the indirect gap well, but provides less accurate values for other band features, including spin-orbit splittings. Our computed Born effective charges and dielectric constants confirm the unusually covalent bonding characteristics of this III-V system.

\end{abstract}

\maketitle

\textit{Introduction:} As microelectronic devices become smaller and more powerful, efficient cooling of their components becomes a critical issue. The transfer of heat from the active component requires materials with very high thermal conductivities. Diamond and graphite, both of which have thermal conductivities $\sim 2000$ Wm$^{-1}$K$^{-1}$, have significant disadvantages: diamond is expensive to produce and graphite is highly anistropic.~\cite{bas_review_properties_angchem_tian2018} A recent first-principles study predicted that BAs has a thermal conductivity comparable to that of diamond and a more favourable thermal expansion coefficient for incorporation in devices.~\cite{bas_dft_highthermal_cond_prl_lindsay2013,bas_dft_gan_thermal_transport_prb_broido2013} Initially, experimental studies failed to confirm this prediction,~\cite{bas_expt_thermal_cond_low_apl_kim2016} as high quality single crystals for measurement are difficult to produce.~\cite{bas_review_properties_angchem_tian2018} Scattering from defects in the single crystals was shown to be a significant factor in the low measured conductivities,~\cite{bas_dft_vacancies_thermal_cond_prb_protik2016,bas_expt_dft_defects_thermtrans_prl_zheng2018} but, after considerable effort in improving crystal growth using the chemical vapour transport method, thermal conductivities of $>1000$ Wm$^{-1}$K$^{-1}$ have been observed,~\cite{bas_expt_high_thermal_cond_science_kang2018,bas_expt_high_thermal_cond_single_crystal_science_li2018,bas_expt_high_thermal_cond_single_crystal_science_tian2018} which are in very close agreement with more recent calculations.~\cite{bas_others_dft_4ph_scattering_lowthermcond_prb_feng2017} The material is therefore of great interest for potential device applications; many of its fundamental properties, however, remain poorly characterised.

BAs stabilises in the zinc blende phase, isostructural with other III-V semiconductors such as GaAs, and has been shown to be $p$-type as-grown.~\cite{bas_expt_ptype_hall_jap_chu1972,bas_dft_bp_bsb_elec-phon_prb_liu2018} Difficulties in growing the material have meant that it has remained relatively obscure until recent years. Early experimental studies indicated a band gap of $\sim 1.5$ eV,~\cite{bas_expt_bandgap_powder_optictransmiss_jelecsoc_ku1966,bas_expt_thinfilm_optical_jelecsoc_chu1974} but whether the gap was direct or indirect was inconclusive. A subsequent measurement on a thin film of BAs indicated an indirect gap of 1.46 eV.~\cite{bas_expt_ptype_electrode_bandgap_jacs_wang2012} Computational studies using wave function methods or density functional theory (DFT) showed that the band structure was similar to that of Si;~\cite{bas_opw_calc_bands_prb_stukel1970,bas_apw_bands_bp_pssb_prasad1989,bas_calc_bands_prb_hart2000} the conduction band minimum (CBM) occurs along the $\Gamma$ to X direction and has p character, in contrast to AlAs, GaAs or InAs, where the s-like CBM is at $\Gamma$. The difference has been attributed to the anomously low energy of the boron p orbitals and strong s-s repulsion,~\cite{bas_calc_bands_prb_hart2000} indicating a high degree of covalency in the bonding between B and As. Moreover, calculations of Born effective charges suggest a reversal of the r\^{o}les of anions and cations in the material,~\cite{bas_dft_gan_thermal_transport_prb_broido2013,zincblende_calc_lattice-dyn_elastic_physicab_pletl1999} in contrast to that expected from formal oxidation states.~\cite{oxidation_states_letter_jphyschemlett_walsh2017,oxidation_states_perspective_nmater_walsh2018} The magnitude of the indirect band gap $E_g^{\mathrm{ind}}$, as well as that of the lowest direct gap (at $\Gamma$) $E_g^{\mathrm{dir}}$ and the band curvatures vary substantially depending on the level of theory applied, with values of $E_g^{\mathrm{ind}}$ ranging from $0.7 - 1.6$ eV.~\cite{bas_expt_dft_defects_thermtrans_prl_zheng2018,bas_opw_calc_bands_prb_stukel1970,bas_apw_bands_bp_pssb_prasad1989,bas_calc_bands_prb_hart2000,bas_dft_bp_bn_pressure_prb_wentzcovitch1987,bas_gw_bn_bp_bands_prb_surh1991,bas_dft_bgaas_binas_props_physicab_chimot2005,bas_dft_alas_gaas_inas_elecprops_compmatersci_ahmed2007,iii-v_comparison_calcs_properties_semicondscitech_anua2013,bas_dft_bp_bsb_elec-phon_prb_liu2018,bas_dft_bands_gw_oneshot_apl_bushick2019,bas_dft_opticalabsorb_bsb_arxiv_ge2019} In a recent study combining theory and experiment, an indirect gap of 1.78 eV was calculated using hybrid DFT,~\cite{bas_expt_dft_hse_defects_bandgap_apl_lyons2018} which agreed well with the value derived from a combination of photoluminescence measurements and computed defect states (1.77 eV).~\cite{bas_expt_dft_hse_defects_bandgap_apl_lyons2018} Other hybrid DFT studies using the same functional, however, report indirect gaps of 1.58,~\cite{bas_dft_opticalabsorb_bsb_arxiv_ge2019} 1.62~\cite{bas_expt_dft_defects_thermtrans_prl_zheng2018} and 1.90 eV.~\cite{bas_dft_defects_hse_apl_chae2018} Further studies using techniques beyond standard DFT, including $GW$, report values from 1.48$-$2.049 eV.~\cite{iii-v_comparison_calcs_properties_semicondscitech_anua2013,bas_dft_elecprops_gaussianorbs_jap_nwigboji2016,bas_gw_bn_bp_bands_prb_surh1991,bas_dft_bgaas_binas_props_physicab_chimot2005,bas_dft_bands_gw_oneshot_apl_bushick2019}

In this Rapid Communication, we employ the relativistic quasiparticle self-consistent $GW$ (QS$GW$) approach~\cite{questaal_qsgw_theory_prl_vanschilfgaarde2006} to compute the band structure and optical properties of BAs. The $GW$ approximation can be used to correct the one-electron eigenvalues obtained from DFT within a many-body quasiparticle framework, including the exchange and correlation effects in a self-energy term dependent on the one-particle Green's function $G$ and the dynamically screened Coulomb interaction $W$. Substantial improvements on DFT energy eigenvalues can be obtained, depending on how the self energy is computed.~\cite{questaal_qsgw_theory_prl_vanschilfgaarde2006,questaal_theory_paper_prb_kotani2007} Previous studies employing the $GW$ approach have been carried out to study the band dispersion of BAs: Surh \textit{et al.},~\cite{bas_gw_bn_bp_bands_prb_surh1991} using an approximation to self-consistent $GW$ that involves a linear interpolation of the DFT eigenvalues, calculated an indirect (direct) gap of 1.6 eV (4.2 eV); Bushick \textit{et al.},~\cite{bas_dft_bands_gw_oneshot_apl_bushick2019} using a single-shot $GW$ approach, where the DFT eigenvalues are used to compute $G$ within a single iteration, computed an indirect (direct) gap of 2.049 eV (4.135 eV); while Chimot \textit{et al.},~\cite{bas_dft_bgaas_binas_props_physicab_chimot2005} also using a single-shot $GW$ approach, reported $E_g^{\mathrm{ind}}=1.87$ eV. The substantial difference between these values indicates their dependency on the underlying DFT calculation (although different lattice parameters used in the calculations and the convergence criteria employed may also play a r\^{o}le). In contrast, the QS$GW$ method solves for the effective potential in a self-consistent manner, resulting in excellent agreement with experiment for a wide range of systems.~\cite{questaal_theory_paper_prb_kotani2007,pbx_qsgw_bands_prb_svane2010,mapi_qsgw_bands_prb_brivio2014,sbchx_qsgw_bands_apl_butler2016,bichi_qsgw_dft_defects_chemmater_ganose2018} Errors introduced tend to be systematic,~\footnote{We note that convergence issues with such computationally expensive techniques may also result in some discrepancies in reported band gaps.} and can be largely accounted for by scaling the QS$GW$ self energy in post-processing runs to compute band structures.~\cite{questaal_calc_bandsplit_prl_chantis2006} Moreover, the errors originate from the lack of ladder diagrams in determining $W$; such effects can be included through solution of the Bethe-Salpeter equation (BSE).~\cite{questaal_theory_paper_prb_kotani2007} By doing so, we expect the approach will give a very accurate energy dispersion, which will be useful for comparison with both experimental studies and computational studies using lower levels of theory.

\textit{Calculations:} To determine the ground state lattice parameter $a$ and electronic structure of BAs, we have used plane-wave DFT as implemented in the \texttt{VASP} code,~\cite{vasp_prb_kresse1993, vasp_prb_kresse1994, vasp_compmatsci_Kresse1996, vasp_prb_kresse1996} utilizing the solids-corrected Perdew-Burke-Ernzerhof (PBEsol) generalised gradient approximation (GGA) exchange-correlation functional~\cite{GGA_pbe_prl_perdew1996, GGA_pbesol_prl_perdew2008} with the projector augmented wave method~\cite{paw_physrevb50_blochl1994} to model the interaction between core and valence electrons (with three valence electrons for B and five for As). The total energy of the BAs zinc blende primitive cell was calculated at a series of constant volumes, using a 500\,eV plane wave cut off and a 12$\times$12$\times$12$ \Gamma$-centred Monkhorst-Pack~\cite{monkhorst_pack_prb_monkhorst1976} \textit{k}-point mesh, which provided convergence in the total energy up to 10$^{-4}$ eV, fitting the resultant energy-volume data to the Murnaghan equation of state. The bulk modulus $B$ and its derivative $B^{\prime}$ were derived using this approach. Spin-orbit interactions were included in the electronic structure calculations,~\cite{spin-orbit_vasp_prb_hobbs2000} while a finer 16$\times$16$\times$16 \textit{k}-point grid was used when computing the density of states (DOS). Furthermore, the calculations were repeated using a hybrid density functional (HSE06~\cite{hse06_functional_jphyschem_heyd2006}), which counters the well-known self-interaction error in DFT. To provide calculated values of physical properties, at the PBEsol level of theory we have computed the high frequency dielectric constant ($\epsilon^{\infty}$), static dielectric constant ($\epsilon^0$), Born effective charges ($Z^*$) and the zone-centre transverse and longitudinal phonon frequencies ($\omega_{\mathrm{TO}}$ and $\omega_{\mathrm{LO}}$, respectively) using density functional perturbation theory, as implemented in \texttt{VASP}.~\cite{optical-prop_paw_rpa_prb_gajdos2006} We have also computed the elastic constants C$_{11}$, C$_{12}$ and C$_{44}$, using the finite displacement approach available in \texttt{VASP}.

The QS$GW$ calculations were performed using the \texttt{Questaal} package,~\cite{questaal_qsgw_theory_prl_vanschilfgaarde2006} which implements DFT and $GW$ within an all-electron linear muffin-tin orbital basis set. Using the relaxed lattice parameter as determined from the plane-wave DFT calculations, the initial wave functions were determined at the PBEsol level of theory, employing the automatically generated augmentation spheres and interstitial Hankel functions as the basis set. The self energy was then obtained using the QS$GW$ formalism employing a $12\times12\times12$ $\Gamma$-centred $k$-point grid for the PBEsol and QS$GW$ calculations which provided convergence in the energy eigenvalues of under 1 meV (a 16$\times$16$\times$16 grid was used to calculate the DOS). The spin-orbit interaction was included perturbatively when computing the eigenvalues using the self energy, as described in Ref.~\onlinecite{mapi_qsgw_bands_prb_brivio2014}. As noted above, QS$GW$ introduces some systematic errors,~\cite{questaal_theory_paper_prb_kotani2007} the most significant of which results in band gap overestimation. This error, however, has been shown to be largely corrected for by scaling the QS$GW$ self energy by 0.8,~\cite{questaal_calc_bandsplit_prl_chantis2006} which is then combined with 0.2 of the PBEsol effective potential when determining band structures. We have followed this approach in the current work, but we have also corrected for the error by including ladder diagrams in $W$ through solution of the BSE,~\cite{questaal_theory_paper_prb_kotani2007} as implemented in \texttt{Questaal}. In this approach, the $G_0$, rather than $G$ is used, which avoids the introduction of unphysical contributions.~\cite{questaal_theory_paper_prb_kotani2007} The procedure consists of a full QS$GW$ calculation, followed by a correction to the $W$ through solution of the BSE. A final calculation of the self energy using the non-interacting $G_0$ is performed in order to compute the band structure.~\cite{questaal_bse_theory_prm_cunningham2018} We compare all three approaches here: the self-consistent method (QS$GW$), the scaled method (sQS$GW$) and the inclusion of vertex corrections (QS$GW$+BSE).

We have calculated the dielectric function $\epsilon(\omega)$ at several levels of theory for comparison. Firstly, we have the summation over empty bands approach implemented in \texttt{VASP}~\cite{optics_sumbands_prb_adolph2001} to compute $\epsilon(\omega)$ at the DFT level of theory. 64 empty bands were found to be sufficient to achieve convergence within three significant figures in this approach. Following the QS$GW$ calculations, we have used the random phase approximation (RPA) to determine $\epsilon(\omega)$, including local field effects via a modified response function, as implemented in \texttt{Questaal}. Finally, going beyond the RPA, we have included excitonic effects through the BSE for the four-point polarisation to compute $\epsilon(\omega)$. We note that, as BAs is an indirect gap system, phonon-assisted absorption is expected to occur at frequencies below the direct optical gap. Such absorption processes are not included in our analysis here but have been calculated elsewhere and shown to be small.~\cite{bas_dft_opticalabsorb_bsb_arxiv_ge2019}

\textit{Results:} Using DFT with the PBEsol functional we determine $a=4.779$\,\AA, while using hybrid DFT with the HSE06 functional yields $a=4.772$\,\AA, both of which are in excellent agreement with the experimental value $a=4.777$\,\AA.~\cite{bas_expt_ptype_hall_jap_chu1972} For the subsequent QS$GW$ calculations, we have imposed the PBEsol-derived $a$ as we use PBEsol-generated wave functions determined with the LMTO basis as the starting point for the self-consistent calculation of the self energy. We discuss below the effect of changing the lattice parameter on the computed band gaps.

\begin{figure*}[ht!]
\centering
\vspace{0.5cm}
\includegraphics*[width=1.0\linewidth]{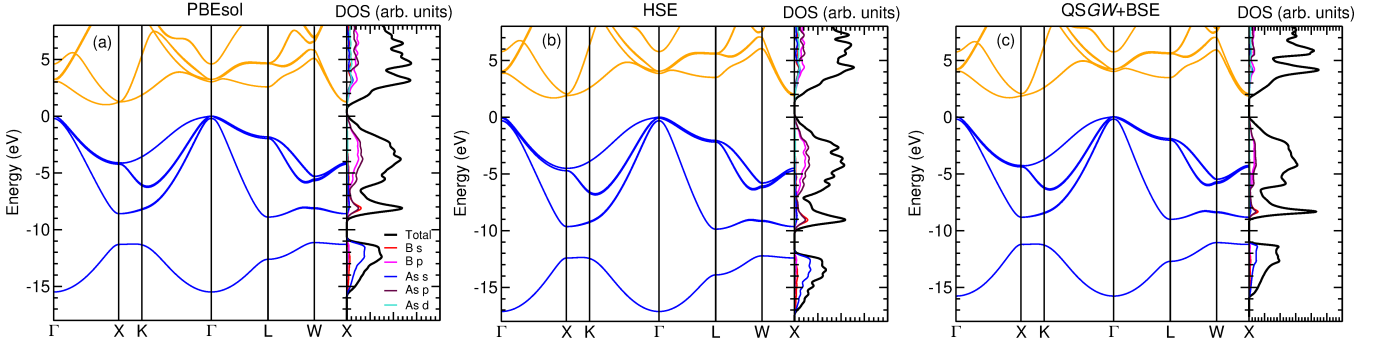}
\caption{Calculated band structure of BAs shown in the zinc blende Brillouin zone determined using (a) DFT with the PBEsol functional; (b) hybrid DFT with the HSE functional; (c) QS$GW$ with ladder diagrams included through solution of the Beth-Salpeter equation (BSE). The zero of the energy scale in each case is the valence band maximum. The density of states (DOS, black line) and partial DOS are also shown in the right-hand figures. The partial DOS is projected onto B s (red line), B p (magenta line), As s (blue line), As p (maroon line) and As d (turquoise line) states.}
\label{fig-bands}
\end{figure*}


Our calculated band structures determined using DFT with the PBEsol functional, hybrid DFT with the HSE06 functional and QS$GW$+BSE are shown in Fig.~\ref{fig-bands}. All three approaches show energy dispersion that is similar to that found in other studies,~\cite{bas_apw_bands_bp_pssb_prasad1989,bas_calc_bands_prb_hart2000,bas_opw_calc_bands_prb_stukel1970,bas_expt_dft_hse_defects_bandgap_apl_lyons2018,bas_dft_bands_gw_oneshot_apl_bushick2019} with an indirect energy gap between the valence band maximum (VBM) at the $\Gamma$ point and the conduction band minimum (CBM) along the $\Gamma$ to X direction, at about 80\% of the distance to the X point and a minimum direct gap at $\Gamma$, but there are differences in particular features such as the band gaps, curvatures and spin-orbit splittings. Indeed, we exclude the band structures determined using QS$GW$ and sQS$GW$ from Fig.~\ref{fig-bands} as they are quite similar to the HSE06 and QS$GW$+BSE results. We instead highlight the differences between the methods in Table~\ref{ener}, where the gaps $E^{\mathrm{ind}}_g$ and $E^{\mathrm{dir}}_g$, spin orbit splittings in the conduction bands at the $\Gamma$ point ($\Delta^{\Gamma,\mathrm{C}}_{\mathrm{SO}}$) and the valence bands at the $\Gamma$, X and L points ($\Delta^{\Gamma,\mathrm{V}}_{\mathrm{SO}}$, $\Delta^{\mathrm{X,V}}_{\mathrm{SO}}$ and $\Delta^{\mathrm{L,V}}_{\mathrm{SO}}$, respectively), $\epsilon^{\infty}$, and the longitudinal (transverse) electron effective mass $m^*_{\mathrm{e},l}$ ($m^*_{\mathrm{e},t}$) at the CBM, the heavy, light and spin-orbit split off hole effective masses ($m^*_{\mathrm{hh}}$, $m^*_{\mathrm{lh}}$ and $m^*_{\mathrm{so}}$, respectively) at the VBM are presented and compared with available values from previous studies.

\begin{table*}[ht]
\caption{Calculated indirect band gap $E^{\mathrm{ind}}_g$, direct band gap $E^{\mathrm{dir}}_g$, valence band (conduction band) spin-orbit split off energy at the $\Gamma$ point $\Delta^{\Gamma,\mathrm{V}}_{\mathrm{SO}}$ ($\Delta^{\Gamma,\mathrm{C}}_{\mathrm{SO}}$), at the X point $\Delta^{\mathrm{X,V}}_{\mathrm{SO}}$ and at the L point $\Delta^{\mathrm{L,V}}_{\mathrm{SO}}$, high-frequency dielectric constant $\epsilon^{\infty}$, longitudinal and transverse electron effective masses ($m^*_{\mathrm{e},l}$ and $m^*_{\mathrm{e},t}$, respectively), heavy hole ($m^*_{\mathrm{hh}}$), light hole ($m^*_{\mathrm{lh}}$) and split-off hole ($m^*_{\mathrm{so}}$) effective masses of BAs. We compare our results obtained using DFT with the PBEsol functional, hybrid DFT with the HSE06 functional, the full QS$GW$ approach and the scaled approach (sQS$GW$) and with the inclusion of ladder diagrams via the Bethe-Salpeter equation (QS$GW$+BSE), with experiment, where available, and previous calculations employing wave function (WF) techniques and DFT, non-local DFT (including hybrid DFT, meta-GGA and the modified Becke Johnson exchange potential) and the $GW$ approach. All energies are given in eV, while the effective masses are given in units of the electronic rest mass.} \centering
\begin{ruledtabular}
\begin{tabular} { c | c c c c c c c c c c }
& \multicolumn{5}{ c }{This Work} & \multicolumn{4}{ c }{Previous Studies} \\
& PBEsol & HSE06 & QS$GW$ & sQS$GW$ & QS$GW$+BSE & WF/DFT & non-local DFT & $GW$ & Experiment \\\hline
$E^{\mathrm{ind}}_g$ & 1.010 & 1.693 & 1.895 & 1.705 & 1.674 & $0.7-1.27$~\cite{bas_opw_calc_bands_prb_stukel1970,bas_apw_bands_bp_pssb_prasad1989,bas_calc_bands_prb_hart2000,bas_dft_props_lda_jphysc_wentzcovitch1986,iii-v_comparison_calcs_properties_semicondscitech_anua2013} & $1.58-1.933$~\cite{bas_dft_opticalabsorb_bsb_arxiv_ge2019,bas_expt_dft_defects_thermtrans_prl_zheng2018,bas_expt_dft_hse_defects_bandgap_apl_lyons2018,bas_dft_defects_hse_apl_chae2018,bas_dft_elecprops_gaussianorbs_jap_nwigboji2016,iii-v_comparison_calcs_properties_semicondscitech_anua2013} & 1.6~\cite{bas_gw_bn_bp_bands_prb_surh1991} & 1.46~\cite{bas_expt_bandgap_powder_optictransmiss_jelecsoc_ku1966,bas_expt_ptype_electrode_bandgap_jacs_wang2012} \\
& & & & & & & & 1.87~\cite{bas_dft_bgaas_binas_props_physicab_chimot2005} & 1.4~\cite{bas_expt_ptype_hall_jap_chu1972} \\
& & & & & & & & 2.049~\cite{bas_dft_bands_gw_oneshot_apl_bushick2019} & 1.77~\cite{bas_expt_dft_hse_defects_bandgap_apl_lyons2018} \\
$E^{\mathrm{dir}}_g$ & 3.032 & 3.846 & 4.216 & 3.966 & 3.990 & 3.56,~\cite{bas_opw_calc_bands_prb_stukel1970} 4.23~\cite{bas_apw_bands_bp_pssb_prasad1989} & 3.05,~\cite{bas_calc_bands_prb_hart2000} 3.7,~\cite{bas_dft_opticalabsorb_bsb_arxiv_ge2019} 3.301~\cite{bas_dft_elecprops_gaussianorbs_jap_nwigboji2016} & 4.2,~\cite{bas_gw_bn_bp_bands_prb_surh1991} 4.135~\cite{bas_dft_bands_gw_oneshot_apl_bushick2019} & \\
$\Delta^{\Gamma,\mathrm{C}}_{\mathrm{SO}}$ & 0.210 & 0.202 & 0.196 & 0.197 & 0.197 & 0.21~\cite{bas_calc_bands_prb_hart2000} & & 0.2,~\cite{bas_gw_bn_bp_bands_prb_surh1991} 0.206~\cite{bas_dft_bands_gw_oneshot_apl_bushick2019} & \\
$\Delta^{\Gamma,\mathrm{V}}_{\mathrm{SO}}$ & 0.207 & 0.316 & 0.211 & 0.210 & 0.207 & 0.33,~\cite{bas_opw_calc_bands_prb_stukel1970} 0.21~\cite{bas_calc_bands_prb_hart2000,bas_dft_bgaas_binas_props_physicab_chimot2005} & 0.23~\cite{bas_expt_dft_hse_defects_bandgap_apl_lyons2018} & 0.22,~\cite{bas_gw_bn_bp_bands_prb_surh1991} 0.206~\cite{bas_dft_bands_gw_oneshot_apl_bushick2019} &  \\
$\Delta^{\mathrm{X,V}}_{\mathrm{SO}}$ & 0.129 & 0.213 & 0.128 & 0.128 & 0.125 & 0.14~\cite{bas_calc_bands_prb_hart2000} & & 0.1,~\cite{bas_gw_bn_bp_bands_prb_surh1991} 0.140~\cite{bas_dft_bands_gw_oneshot_apl_bushick2019} & \\
$\Delta^{\mathrm{L,V}}_{\mathrm{SO}}$ & 0.144 & 0.145 & 0.144 & 0.144 & 0.141 & 0.17~\cite{bas_calc_bands_prb_hart2000} & & 0.1,~\cite{bas_gw_bn_bp_bands_prb_surh1991} 0.148~\cite{bas_dft_bands_gw_oneshot_apl_bushick2019} & \\
$\epsilon^{\infty}$ & 9.909 & 7.995 & 7.384 & 7.690 & 9.881 & 9.8,~\cite{bas_dft_bgaas_binas_props_physicab_chimot2005} 8.1779~\cite{iii-v_comparison_calcs_properties_semicondscitech_anua2013} & & 9.02~\cite{bas_dft_bands_gw_oneshot_apl_bushick2019} & \\
$m^*_{\mathrm{e},l}$ & 1.128 & 1.016 & 1.028 & 1.036 & 1.004 & 1.2,~\cite{bas_opw_calc_bands_prb_stukel1970} 1.099~\cite{bas_dft_bp_bsb_elec-phon_prb_liu2018} & 1.16,~\cite{bas_expt_dft_hse_defects_bandgap_apl_lyons2018} 1.015~\cite{bas_dft_elecprops_gaussianorbs_jap_nwigboji2016} & 1.093~\cite{bas_dft_bands_gw_oneshot_apl_bushick2019} & \\
$m^*_{\mathrm{e},t}$ & 0.198 & 0.198 & 0.199 & 0.201 & 0.200 & & 0.32,~\cite{bas_expt_dft_hse_defects_bandgap_apl_lyons2018} 0.232~\cite{bas_dft_elecprops_gaussianorbs_jap_nwigboji2016} & 0.239~\cite{bas_dft_bands_gw_oneshot_apl_bushick2019} & \\
$m^*_{\mathrm{hh}}$ & 0.420 & 0.374 & 0.605 & 0.610 & 0.624 & 0.51,~\cite{bas_opw_calc_bands_prb_stukel1970} 0.402,~\cite{bas_dft_bp_bsb_elec-phon_prb_liu2018} 0.403~\cite{bas_dft_bgaas_binas_props_physicab_chimot2005} & 0.64,~\cite{bas_expt_dft_hse_defects_bandgap_apl_lyons2018} 0.355~\cite{bas_dft_elecprops_gaussianorbs_jap_nwigboji2016} & 0.635~\cite{bas_dft_bands_gw_oneshot_apl_bushick2019} & \\
$m^*_{\mathrm{lh}}$ & 0.172 & 0.188 & 0.129 & 0.130 & 0.129 & 0.20,~\cite{bas_opw_calc_bands_prb_stukel1970} 0.182,~\cite{bas_dft_bp_bsb_elec-phon_prb_liu2018} 0.193~\cite{bas_dft_bgaas_binas_props_physicab_chimot2005} & 0.25,~\cite{bas_expt_dft_hse_defects_bandgap_apl_lyons2018} 0.107~\cite{bas_dft_elecprops_gaussianorbs_jap_nwigboji2016} & 0.192~\cite{bas_dft_bands_gw_oneshot_apl_bushick2019} & \\
$m^*_{\mathrm{so}}$ & 0.221 & 0.225 & 0.264 & 0.263 & 0.263 & & 0.26~\cite{bas_expt_dft_hse_defects_bandgap_apl_lyons2018} & 0.192~\cite{bas_dft_bands_gw_oneshot_apl_bushick2019} & \\
\end{tabular}
\end{ruledtabular}
\label{ener}
\end{table*}

In all three cases presented in Fig.~\ref{fig-bands}, the total DOS and partial DOS indicates that the upper valence band consists of mixing between B and As p orbitals, while the lower conduction bands are dominated by B p states, demonstrating the high degree of covalency in this system. This result is in contrast to other III-V arsenides, but consistent with the study of Hart and Zunger.~\cite{bas_calc_bands_prb_hart2000} From our DFT calculation using the PBEsol functional we find that $E^{\mathrm{ind}}_g=1.010$ eV, which is lower than that found in other studies using similar GGA functionals~\cite{bas_dft_alas_gaas_inas_elecprops_compmatersci_ahmed2007,bas_dft_bp_bsb_elastic_elec_props_pssb_meradji2004,iii-v_comparison_calcs_properties_semicondscitech_anua2013,bas_dft_opticalabsorb_bsb_arxiv_ge2019,bas_expt_dft_defects_thermtrans_prl_zheng2018} by about 0.2 eV. We note, however, that those studies do not include the spin-orbit interaction (SOI); we attribute the discrepancies to differences in the GGA functionals used and the inclusion of the SOI. Indeed, we find that, in all our calculations, including the SOI reduces $E^{\mathrm{ind}}_g$ by 0.07 eV and $E^{\mathrm{dir}}_g$ by 0.2 eV (apart from the HSE06 case, where $E^{\mathrm{ind}}_g$ is reduced by 0.14 eV and $E^{\mathrm{dir}}_g$ by 0.27 eV).

Using the HSE06 functional increases the band gaps, as expected. Our value of $E^{\mathrm{ind}}_g=1.693$ eV using HSE06 functional is larger than those of Ge \textit{et al.}~\cite{bas_dft_opticalabsorb_bsb_arxiv_ge2019} (1.58 eV) and Zheng \textit{et al.}~\cite{bas_expt_dft_defects_thermtrans_prl_zheng2018} (1.62 eV), who did not include the SOI. Zheng \textit{et al.} employed lower convergence criteria and Ge \textit{et al.} used a different approach, in both cases finding lattice parameters larger than our calculated value using HSE06 (4.80 and 4.818 \,\AA\, \textit{versus} 4.772 \,\AA), which may account for the lower computed band gaps. Chae \textit{et al.}~\cite{bas_dft_defects_hse_apl_chae2018} determined $E^{\mathrm{ind}}_g=1.90$ eV using the HSE06 functional without the SOI and with a lower plane-wave cut off (400 eV); their result is reasonably close to our value excluding the SOI (1.833 eV). We note that Lyons \textit{et al.}~\cite{bas_expt_dft_hse_defects_bandgap_apl_lyons2018} did include the SOI and calculated $E^{\mathrm{ind}}_g=1.78$ eV, using an approach very similar to ours (but also with the lower cut off of 400 eV). They included 3d states as valence states in As, which they determined to lie $\sim40$ eV below the VBM. It is not clear, therefore, what advantage is gained by their inclusion; we note that the pseudopotential that excludes the 3d states from the core is in much less use than the standard, five valence electron pseudopotential, and may therefore be less robust.

\begin{figure}[ht!]
\centering
\vspace{0.5cm}
\includegraphics*[width=1.0\linewidth]{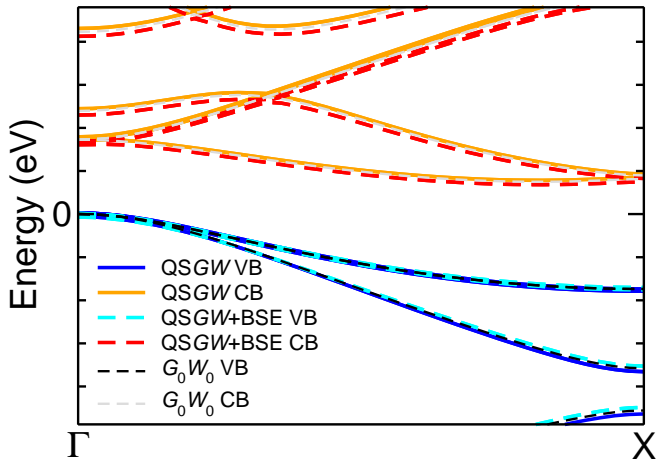}
\caption{Comparison of the calculated bands along the $\Gamma$ to X direction determined using the QS$GW$ approach and via the inclusion of ladder diagrams by solution of the Bethe-Salpeter equation (BSE). The valence bands (VBs) are indicated by continuous blue (broken turquoise) lines and the conduction bands (CBs) by continuous orange (broken red) lines for the QS$GW$ (QS$GW$+BSE) case. Thinner broken black (grey) lines indicate the VBs (CBs) as determined using the single-shot $GW$ ($G_0W_0$) approach. The zero of the energy scale in each case is the valence band maximum.}
\label{fig-bse}
\end{figure}

Including quasiparticle energies with the QS$GW$ approach increases the band gap significantly.~\footnote{Using the LMTO basis, with the PBEsol functional we determine $E^{\mathrm{ind}}_g=1.015$ eV, in very good agreement with our plane-wave DFT-PBEsol calculation.} We first note that, with a single-shot $GW$ calculation based on the PBEsol-derived wave functions, we determine $E^{\mathrm{ind}}_g=1.805$ eV and $E^{\mathrm{dir}}_g=4.118$ eV (including the SOI, as we do for all the QS$GW$ calculations reported below). As expected, these results differ from previous one-shot $GW$ calculations,~\cite{bas_gw_bn_bp_bands_prb_surh1991,bas_dft_bgaas_binas_props_physicab_chimot2005,bas_dft_bands_gw_oneshot_apl_bushick2019} which were based on the local density approximation (LDA). The self-consistent prodecure requires three further iterations, resulting in wider gaps of $E^{\mathrm{ind}}_g=1.895$ eV and $E^{\mathrm{dir}}_g=4.216$ eV. These values are overestimates, due to systematic errors in the approach used,~\cite{questaal_theory_paper_prb_kotani2007} but can be largely corrected for using the QS$GW$+BSE method, which reduces the band gaps by over 0.2 eV (see Table~\ref{ener}). In fact, the conduction bands are shifted down consistently, as can be seen in Fig.~\ref{fig-bse}, where we also include the single-shot $GW$ bands for comparison (as the QS$GW$+BSE procedure includes the $G_0$, rather than $G$ determined in the QS$GW$ method).~\cite{questaal_bse_theory_prm_cunningham2018} Including a simple scaling of the self energy (sQS$GW$) results in band gaps that are within a few tens of meV to the full QS$GW$+BSE calculation, but the derived optical properties have significant differences (see below). It is worth noting that our calculated fundamental gap agrees well with a previous calculation which used the modified Becke-Johnson exchange potential~\cite{tb-mbj_exchange_test_prl_tran2009}, excluding the SOI (1.73 eV).~\cite{iii-v_comparison_calcs_properties_semicondscitech_anua2013} 

Using the QS$GW$+BSE approach, the resulting $E^{\mathrm{ind}}_g=1.674$ eV is larger than that measured in early experimental studies (about 1.5 eV)~\cite{bas_expt_bandgap_powder_optictransmiss_jelecsoc_ku1966,bas_expt_ptype_hall_jap_chu1972} and in a more recent study using thin films (1.46 eV),~\cite{bas_expt_ptype_electrode_bandgap_jacs_wang2012} but crystal quality, strain in the case of the thin film, as well as temperature effects may account for the discrepancies. A value of 1.77 eV, derived from photoluminescence measurements,~\cite{bas_expt_dft_hse_defects_bandgap_apl_lyons2018} depends strongly on computed defect levels, which are notoriously difficult to determine with accuracies better than $\sim0.1$ eV in many cases.~\cite{in2o3_sno2_zno_prm_buckeridge2018} Our result would be compatible with the experimental measurement if slightly shallower defect states exist in the material. We note that, from our results the HSE06 functional reproduces the fundamental gap quite well. Other features, however, such as $E^{\mathrm{dir}}_g$, $\Delta^{\Gamma,\mathrm{V}}_{\mathrm{SO}}$ and $\Delta^{\mathrm{X,V}}_{\mathrm{SO}}$ disagree substantially with those calculated using QS$GW$, sQS$GW$ and QS$GW$+BSE (and indeed DFT-PBEsol). The $a$ used in the HSE06 calculations differs to that in the PBEsol-derived QS$GW$ approaches; if we instead use the $a$ determined using DFT-PBEsol in the HSE06 calculation, we find only a very slight change in the band gaps: by 4 meV for $E^{\mathrm{ind}}_g$ and by -6 meV for $E^{\mathrm{dir}}_g$.

We have calculated the carrier effective masses at the band edges (see Table~\ref{ener}). For the case of the DFT and hybrid DFT calculations, we have applied quadratic fits to the bands within 0.001 eV of the appropriate band extremum, along and perpendicular to the $\Gamma$ to X direction for the case of electrons, and along the $\Gamma$ to X, L and K directions for holes. For the hole effective masses, where the bands are non-spherical, we took an average of the values obtained. Our results are in reasonable agreement with other calculations using similar approaches (DFT and hybrid DFT, see Table~\ref{ener}).~\cite{bas_dft_bp_bsb_elec-phon_prb_liu2018,bas_opw_calc_bands_prb_stukel1970,bas_dft_bgaas_binas_props_physicab_chimot2005,bas_expt_dft_hse_defects_bandgap_apl_lyons2018,bas_dft_elecprops_gaussianorbs_jap_nwigboji2016} For the QS$GW$ calculations, however, we took advantage of the \texttt{Questaal} code, which allows one to fit a quadratic function to a set of points forming an icosohedron about the band edges and derive the effective mass tensor. We therefore expect this approach to provide more accurate results than simply fitting to the bands along high symmetry directions (although for electrons the two approaches should be equivalent). The resulting hole effective masses were taken to be the averages of the diagonal mass tensor components. In all cases, the agreement between the results obtained using the different approaches is reasonable. We find that, similar to other studies,~\cite{bas_dft_bands_gw_oneshot_apl_bushick2019} holes are lighter than electrons and should therefore be more mobile.



\begin{figure}[ht!]
\centering
\vspace{0.5cm}
\includegraphics*[width=1.0\linewidth]{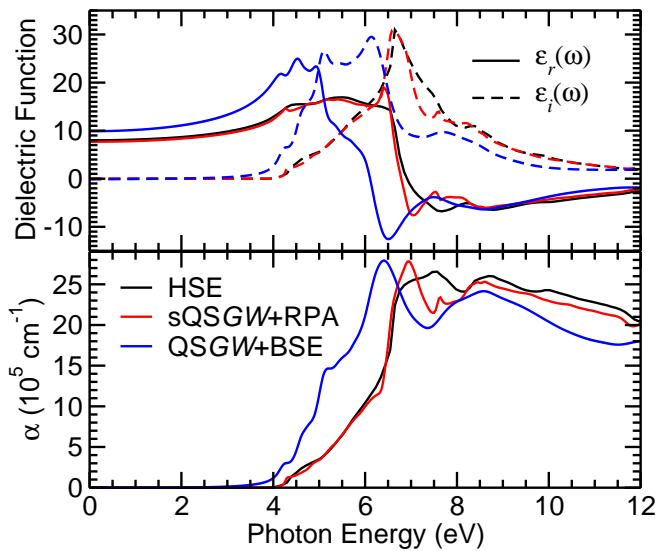}
\caption{The calculated real, $\epsilon_r(\omega)$, and imaginary, $\epsilon_i(\omega)$, parts of the dielectric function (top panel) determined using hybrid DFT with the HSE functional (black lines), scaled QS$GW$ (see text) within the random phase approximation (RPA, red lines) and via the inclusion of ladder diagrams by solution of the Bethe-Salpeter equation (BSE, blue lines). $\epsilon_r(\omega)$ ($\epsilon_i(\omega)$) is indicated by a solid (broken) line. The bottom panel shows the derived optical absorption, $\alpha(\omega)$.}
\label{fig-op}
\end{figure}


The calculated values of $\epsilon^{\infty}$ given in Table~\ref{ener} indicate the importance of including excitonic effects. $\epsilon^{\infty}$ decreases approximately with increasing band gap, from 9.909 (PBEsol) to 7.384 (QS$GW$), but increases substantially to 9.881, very close to the PBEsol value, when the QS$GW$+BSE approach is used. We have calculated the dielectric functional $\epsilon(\omega)=\epsilon_r(\omega)+\epsilon_i(\omega)$ using the summation over empty bands approach with the HSE06 functional, the sQS$GW$+RPA and the QS$GW$+BSE methods, as explained above. The results are shown in Fig.~\ref{fig-op}, with the derived optical absorption $\alpha(\omega)$ shown in the bottom panel. The HSE06 and sQS$GW$+RPA results are very similar, apart from an additional peak at about $\hbar\omega=7.5$ eV in $\epsilon_i(\omega)$, which may be due to some small differences in the band splittings between the HSE06 and sQSGW band structures. Including excitonic effects, however, substantially changes the optical properties. An onset in absorption is observed just below the direct gap (3.990 eV), a large additional peak is observed in $\epsilon_i(\omega)$ at about $\hbar\omega=5$ eV and the next peak is shifted down by about 0.5 eV, as a result of the shifting downwards of the conduction bands (see Fig.~\ref{fig-bse}). As BAs is an indirect gap system, phonon-assisted transitions will occur below the direct gap, which we do not inculde in our analysis. Their effect, however, is expected to be minor.~\cite{bas_dft_opticalabsorb_bsb_arxiv_ge2019} We note that the result of including the excitonic effects is similar to that seen in Ref.~\onlinecite{bas_dft_bands_gw_oneshot_apl_bushick2019}.

\begin{table*}[ht]
\caption{Calculated bulk modulus $B_0$, its derivative $B^{\prime}_0$, elastic constants C$_{11}$, C$_{12}$ and C$_{44}$, zone-centre longitudinal (transverse) optical phonon frequency $\omega_{\mathrm{LO}}$ ($\omega_{\mathrm{TO}}$), Born effective charges $Z^*$ and static dielectric constant $\epsilon^0$ of BAs, as computed using DFT with the PBEsol functional.} \centering
\begin{ruledtabular}
\begin{tabular} { c c c c c c c c c c }
$B_0$ (GPa) & $B^{\prime}_0$ & C$_{11}$ (GPa) & C$_{12}$ (GPa) & C$_{44}$ (GPa) & $\omega_{\mathrm{LO}}$ (cm$^{-1}$) & $\omega_{\mathrm{TO}}$ (cm$^{-1}$) & $Z^*_{\mathrm{B}}$ & $Z^*_{\mathrm{As}}$ & $\epsilon^0$\\ \hline
139.0 & 3.99 & 275.8 & 73.3 & 168.7 & 692.4 & 696.0 & -0.509 & 0.503 & 10.010 \\ \hline
\end{tabular}
\end{ruledtabular}
\label{bulk_ps}
\end{table*}

Finally, we have computed the elastic, dielectric and lattice dynamical properties of BAs using DFT with the PBEsol functional. The elastic constants and bulk modulus are in good agreement with previous calculations,~\cite{iii-v_comparison_calcs_properties_semicondscitech_anua2013,zincblende_calc_lattice-dyn_elastic_physicab_pletl1999,bas_dft_bgaas_binas_props_physicab_chimot2005} as are the zone centre optical phonon frequencies.~\cite{bas_dft_gan_thermal_transport_prb_broido2013,bas_dft_opticalabsorb_bsb_arxiv_ge2019,bas_dft_vacancies_thermal_cond_prb_protik2016,bas_dft_bgaas_binas_props_physicab_chimot2005,zincblende_calc_lattice-dyn_elastic_physicab_pletl1999} $\epsilon^0$, at 10.010, is very close to $\epsilon^{\infty}$ (9.909, see Table~\ref{ener}); consequently we have a lattice contribution to the dielectric constant of 0.101, indicating the high degree of covalency in this system. This result is in good agreement with Bushick \textit{et al.}~\cite{bas_dft_bands_gw_oneshot_apl_bushick2019} The $Z^*$ indicate a surprising r\^{o}le reversal of the cation (B) and anion (As), with a negative value of $Z^*_{\mathrm{B}}=-0.509$. Combined with our calculated refractive index $n=3.15$, this result is in good agreement with previous studies that report the reduced charge, $Z^*/n$,~\cite{bas_dft_gan_thermal_transport_prb_broido2013,zincblende_calc_lattice-dyn_elastic_physicab_pletl1999} and is consistent with the bonding description given by Hart and Zunger.~\cite{bas_calc_bands_prb_hart2000}


\textit{Summary:} We report the band structure and optical properties of BAs derived using the QS$GW$+BSE approach, determing an indirect gap of 1.674 eV and direct gap of 3.990 eV. We have demonstrated the importance of including self consistency in the $GW$ calculation and the inclusion of excitonic effects when determining the quasiparticle spectra by comparing our results to those we have obtained using DFT and hybrid DFT, as well as other previous calculations and experiment. Our computed effective masses agree with previous studies, as do the calculated bulk elastic, dielectric and lattice dynamical properties obtained using the PBEsol functional within DFT. The results indicate the unusually strong covalent bonding properties in this III-V arsenide and a swapping of the anion and cation r\^{o}les, consistent with previous studies.

\section*{Acknowledgment}

The authors acknowledge the use of the UCL Legion and Grace High Performance Computing Facilities (Legion@UCL and Grace@UCL) and associated support services, and the ARCHER supercomputer through membership of the UK's HPC Materials Chemistry Consortium, which is funded by EPSRC grant EP/L000202, in the completion of this work. D. O. S. acknowledges membership of the Materials Design Network. We would like to thank A. J. Jackson, B. A. D. Williamson and C. N. Savory for useful discussions.

\bibliographystyle{aip} 


\end{document}